# Coronavirus research before 2020 is more relevant than ever, especially when interpreted for COVID-19

Mike Thelwall, Statistical Cybermetrics Research Group, University of Wolverhampton, UK.

The speed with which biomedical researchers were able to identify and characterise COVID-19 was clearly due to prior research with other coronaviruses. Early epidemiological comparisons with two previous coronaviruses, Severe Acute Respiratory Syndrome (SARS) and Middle East Respiratory Syndrome (MERS), also made it easier to predict COVID-19's likely spread and lethality. This article assesses whether academic interest in prior coronavirus research has translated into interest in the primary source material, using Mendeley reader counts for early academic impact evidence. The results confirm that SARS and MERS research 2008-2017 experienced anomalously high increases in Mendeley readers in April-May 2020. Nevertheless, studies learning COVID-19 lessons from SARS and MERS or using them as a benchmark for COVID-19 have generated much more academic interest than primary studies of SARS or MERS. Thus, research that interprets prior relevant research for new diseases when they are discovered seems to be particularly important to help researchers to understand its implications in the new context.
**Keywords**: COVID-19; SARS; MERS; Mendeley; Altmetrics; Readers.

## 1 Introduction

The COVID-19 pandemic in 2020 was recognised because scientists already knew about coronaviruses. The origins of coronaviruses were also already known (zoonotic with specific animal carriers). Moreover, an understanding of virus mutations had led to an expectation that new coronaviruses could emerge and that their virulence could differ from those already found. Thus, whilst the virulence of COVID-19 and timing of its occurrence could not be predicted in advance, its emergence was a recognised possibility.

In addition, prior coronavirus research had identified a set of symptoms from previous outbreaks, tested a range of treatments, experimented with vaccines, and implemented preventative measures. Thus, biomedical and public health investigations of COVID-19 had a body of prior coronavirus research to draw upon. Assessing the extent to which COVID-19 differs from prior diseases might help speed new biomedical and public health research, for example. This is made explicit in some papers, such as "Repurposing antivirals as potential treatments for SARS-CoV-2: From SARS to COVID-19" (Gómez-Ríos, López-Agudelo, & Ramírez-Malule, 2020). It seems likely that this is a general trend, so older coronavirus research will be attracting substantial new attention in 2020, but evidence is needed to confirm this.

There are currently three known coronaviruses diseases that can have a serious impact on humans. Other coronaviruses are mild in humans or only infect some species of animals.

- **SARS** (Severe Acute Respiratory Syndrome) is caused by the coronavirus SARS-CoV (also known as SARS-CoV-1, SARSr-CoV). It was first identified in 2003, and there has been no outbreak since then. 8437 people have been reported infected, with an 11% death rate (https://www.who.int/csr/sars/country/2003_07_11/en/).
- **MERS** (Middle East Respiratory Syndrome) is caused by the MERS coronavirus MERS-CoV. It was first identified in Saudi Arabia in 2012 and by January 2020, 2500 people had been reported infected, with a 35% death rate (http://www.emro.who.int/health-topics/mers-cov/mers-outbreaks.html)

- **COVID-19** is caused by the coronavirus SARS-CoV-2 and emerged in December 2019. At the time of writing, it had infected many more people than the previous two coronaviruses, was more infectious for human-to-human transmission, had a lower death rate but much higher death toll. It has previously been called 2019-nCoV and 2019/2020 novel coronavirus.

Despite the above-mentioned likelihood of prior coronavirus research being more useful in 2020, there is no evidence yet to check whether interest in research specific to SARS and MERS has increased due to COVID-19. A positive result – even though highly expected - would empirically validate the importance of ongoing research into diseases related to potential pandemics (e.g., coronaviruses, ebolaviruses, Flaviviridae viruses). This article addresses this issue and compares the current academic impact of COVID-19 research with prior coronavirus research to assess their current relative importance. It is not clear whether older coronavirus research would be more impactful. This seems like a possibility because it may be more foundational, and higher quality due to more time to plan and execute. Conversely, research focusing on COVID-19 may be more relevant to the 2020 pandemic. The research questions are as follows.

- RQ1: Is SARS and MERS research from before 2020 more cited in 2020 than expected for its age?
- RQ2: Is SARS and MERS research from before 2020 having more scholarly impact than 2020 COVID-19 research?

## 2 Background: Bibliometric studies of coronaviruses

Some bibliometric studies have investigated the influence of coronavirus research, mostly characterising the number and type of publications indexed in relevant scholarly databases. Since coronaviruses have many variants and could be the primary focus of a paper or less central to a research project, each study has operationalised its sample in different ways and there is not a single agreed method. All seem to have been produced in 2020 in the context of COVID-19.

A range of studies have shown that there is a rapid rate of COVID-19 research publishing and that both MERS and SARS are relevant to this emerging set. In detail, and discussed chronologically, the specific findings are as follows. One study found 8732 articles and 1028 reviews with the title term "coronavirus" by February 9, 2020 in the Web of Science. The Journal of Virology (9%) was the single most common source and both SARS and MERS were identified as relevant keywords (Tao, Zhou, Yao, et al., 2020). By February 29, 2020, 183 publications matching the query, "COVID-19" were indexed in PubMed, a third of which reported original research (Lou, Tian, Niu, et al., 2020). In PubMed and the World Health Organisation (WHO) COVID-19 research database, there had been 564 observational or interventional investigations about COVID-19 by March 18, 2020 (Chahrour, Assi, Bejjani, et al., 2020). An investigation of Web of Science (WoS) publications matching a set of COVID-19 queries on 1 April 2020 found keyword related to both MERS and SARS to be associated, suggesting that early research had often made connections with the two prior diseases (Hossain, 2020). By 7 April, the COVID-19 coverage of a range of scholarly databases had been increasing at an increasing rate since January 2020 (Torres-Salinas, 2020). On 9 April, 12,109 papers had been indexed by Scopus matching the query "coronavirus*", with sudden increases associated with each of SARS, MERS and COVID-19 (Haghani, Bliemer, Goerlandt, & Li, 2020; see also: Danesh & GhaviDel, 2020). A collection of 2958 articles and 2797 preprints from Scopus, arXiv, bioRxiv, and medRxiv by April 23, 2020 was created by a set of inclusive queries, with unspecified manual checking afterwards (Latif, Usman, Manzoor, et al., 2020). Topic modelling applied to this dataset was used extracted sets of ten topics for different slices of the data, but none included SARS or MERS. Scopus queries for “COVID-19” in titles, abstracts or keywords two days later (April 25) found 3,513 documents and a keyword-based topic visualisation included MERS (Hamidah, Sriyono, & Hudha, 2020). Also on April 25, the number of COVID-19 documents indexed by PubMed was experiencing exponential-like growth (Kambhampati, Vaishya, & Vaish, 2020). A similar investigation with a wider range of queries found both SARS and MERS represented in a topic map (Dehghanbanadaki, Seif, Vahidi, et al., 2020).

One prior bibliometric study has compared SARS, MERS and COVID-19 papers until March 25, 2020. Similar document types were published for each disease, both SARS and MERS research volume had decreased over time, and COVID-19 research was more cited (higher field normalised citation counts). This paper adopted an inclusive search strategy, finding 7,272 SARS documents and 2,199 MERS documents (Hu, Chen, Wang, et al., 2020). Thus, this analysis may be dominated by studies that are related to the three diseases without being primarily about them.

Some research has compared bibliometric trends for a variety of diseases, including coronaviruses. A comparison of SARS, MERS, Avian Flu, Ebola, HIV/AIDS, Hepatitis B & C, Flu, and Swine Flu found that short-lasting epidemics were uniquely associated with rapid increases and declines in both publication volumes and citation rates around the critical years (Kagan, Moran-Gilad, & Fire, 2020). Another study compared COVID-19 with SARS, Ebola, Avian Flu (H1N1), and Zika publications in WoS by 9 April 2020 (adding papers from PubMed and the Chinese National Knowledge Infrastructure for COVID-19). It found that all four prior epidemics were associated with rapid increases in publication volumes, with slower declines. Research into all diseases covered a wide range of subject areas (Zhang, Zhao, Sun, Huang, & Glänzel, 2020).

# 3 Methods

The research design was to gather studies about coronaviruses from before 2020 and compare their impact with that of studies published in 2020. There is a large curated relevant dataset, the COVID-19 Open Research Dataset (CORD-19), which is a collection of papers designed for data mining and scientometrics but takes a broad approach by including some non-COVID-19 publications (Colavizza, Costas, Traag, Van Eck, Van Leeuwen, & Waltman, 2020). This was not used because of its broad remit. The scholarly database Dimensions was chosen in preference to the Web of Science, PubMed or Scopus for its rapid indexing of academic documents and wider coverage of COVID-19 than other scholarly indexes (Torres-Salinas, 2020; Kousha & Thelwall, 2020). For the basic sample, Dimensions was searched for documents matching “coronavirus” weekly from 21 March 2020 to 30 May 2020, using the query below. This single term was chosen, rather than set of coronavirus-related keywords and phrases, to give a narrow focus on the virus. The earliest result was from 2008, which seems to be an API limitation since the web version has results from 1950.

search publications for "coronavirus" return publications [basics + extras]

Mendeley was queried for each document matching the above query in order to count the number of registered readers of the document. Reader counts were checked each week, immediately after the Dimensions queries. Mendeley readers have moderate or strong correlations with citation counts in all or almost all academic fields (Thelwall, 2017b) and recent scholarly articles usually have at least one Mendeley reader (Zahedi, Costas, & Wouters, 2014). Early Mendeley reader counts have a high correlation with later Scopus citation counts (Thelwall, 2018) but appear about a year before them (Maflahi & Thelwall, 2018; Thelwall, 2017a), so they are preferable to citation counts as an indicator of academic impact. This is also true for COVID-19 research (Kousha & Thelwall, 2020). Most people registering documents in Mendeley are academics or PhD students, although there are also some master’s students, librarians and other professionals (Mohammadi, Thelwall, Haustein, & Larivière, 2015). People usually register an article in Mendeley because they have read it or intend to read it (Mohammadi, Thelwall, & Kousha, 2016) so all evidence points to it being a citation-like academic impact indicator, with a small element of educational impact.

Mendeley documents were searched for using an author/title query and a separate DOI query, with the results combined to identify readers of all variants of a document in Mendeley (Zahedi, Haustein, & Bowman, 2014). The Mendeley reader counts were not field normalised because the documents fit within a relatively narrow topic and trends over time are clearer with the raw reader count data.

Four subsets were extracted to identify the influence of different types of research. The first three sets of documents were based on the inclusion of a human coronavirus-related keyword in the title. The fourth encapsulated any mention of coronaviruses generally or a specific human coronavirus in article titles. Although an article can be about these topics without containing a disease or virus name in the title, for example by being published before a formal name was assigned (e.g., "A pneumonia outbreak associated with a new coronavirus of probable bat origin"), or research targeting an aspect of the disease without needing to specify it (e.g., "The psychological impact of quarantine and how to reduce it: rapid review of the evidence" and "First respiratory transmitted food borne outbreak?" from 2020), this method seemed to be effective at eliminating peripherally relevant papers. For example, many papers in the complete set of matches of the original Dimensions coronavirus query were about other viruses or about viruses in general but mentioned coronaviruses as an example or as part of a list (e.g., "Fighting misconceptions to improve compliance with influenza vaccination among health care workers").

- **MERS**: Journal articles with "MERS", "MERS-CoV" or "Middle East Respiratory Syndrome" in their titles.
- **SARS**: Journal articles with "SARS", "SARSr-CoV", "SARS-CoV-1" or "SARS-CoV", or "Severe Acute Respiratory Syndrome" in their titles. The results from 2020 were manually checked to remove false matches that were mentions of COVID-19 before it had been named.
- **COVID-19**: Journal articles containing "COVID-19", "COVID19", "COVID2019", "SARS-CoV-2", "2019-nCoV", "2019 coronavirus", "coronavirus disease 2019", or "Wuhan", in their titles. The inclusion of "Wuhan" did not generate false matches because the original dataset was captured with a coronavirus query.
- **Coronavirus**: Journal articles containing any of the above or the word "Coronavirus" in their titles. This encapsulates the three human coronaviruses and the generic name for the virus family, but not names for other coronaviruses.

Documents that did not match the Dimensions classification for journal articles were excluded. Some of the journal articles may have been news items, editorials, letters, short articles or reviews rather than standard journal articles. These were retained because short form contributions seem to play an important role in infectious disease research (Kousha & Thelwall, 2020).

The rate of increase for each subset was calculated with the percentage increase in average Mendeley readers over about two months: from the first date checked in April to the last date in May 2020. Although the start date could have been earlier, at 21 March 2020, there was a higher rate of increase at the end of March. April was chosen as the starting point as a conservative step, in case the end of March increase was due to a technical cause.

# 4 Results

As captured by Dimensions, the volume of research about SARS has slowly decreased since 2011 (Figure 1), with a projected decrease for 2020, even though the 2020 data contains only a quarter of a year (articles recorded in Mendeley on 21 March 2020). In contrast, the volume of research about MERS increased to 2015/16, then decreased, even allowing for the 2020 data containing only a quarter of a year.

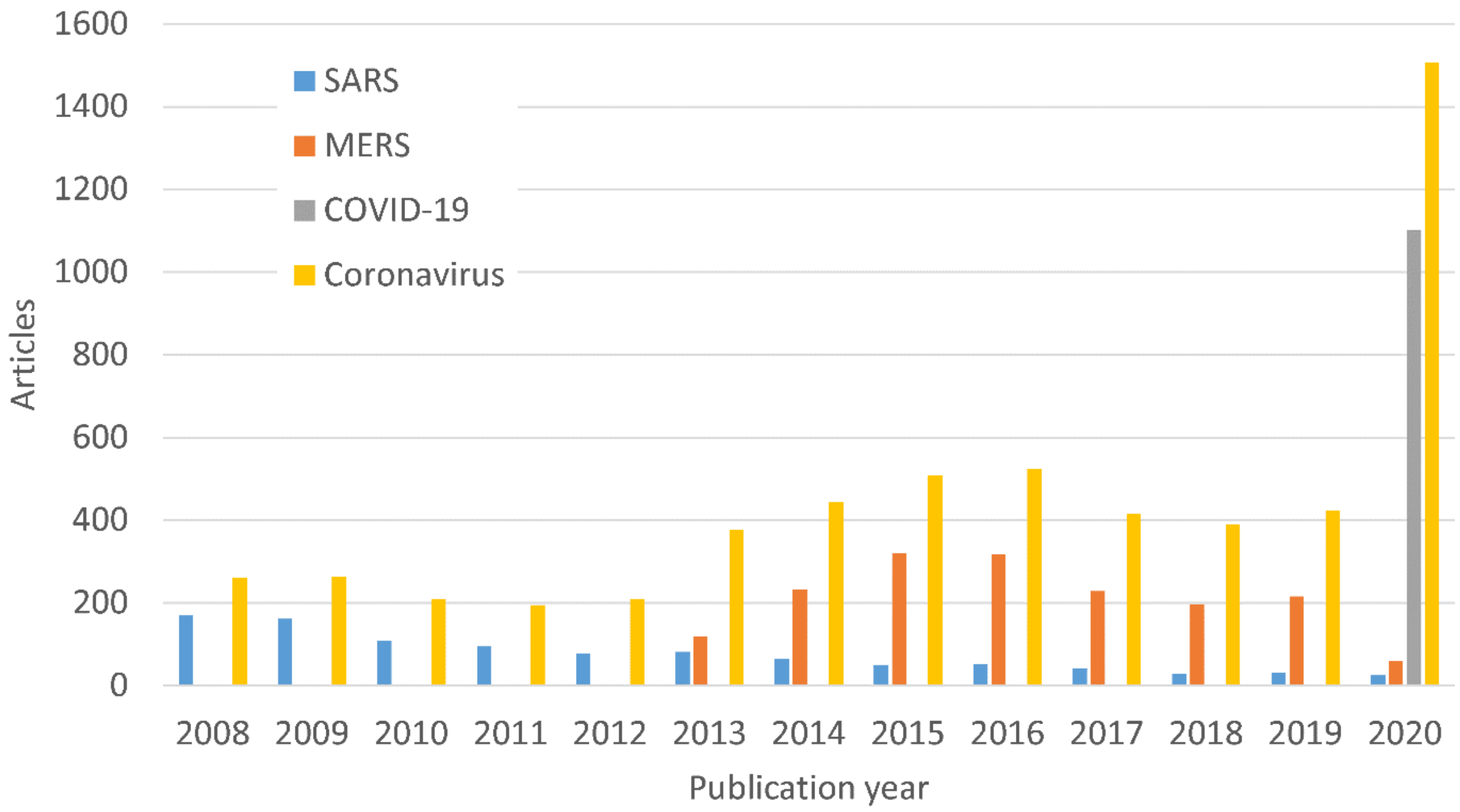

Figure 1. The number of journal articles matching Dimensions "coronavirus" queries and with a title containing "SARS", "MERS" or "Coronavirus", and with a Mendeley record on 21 March 2020. "SARS" documents exclude those with "SARS-" in the title.

The average readership for the three sets of articles is very approximately constant, irrespective of year, for all sets, except with a substantial increase in 2020. Other factors being equal, older articles should be more cited and have more readers because interest should accrue over time. Thus, the relatively static average numbers of readers per article until 2019 suggests a moderate tendency for newer articles to be more read in all three categories. In addition, articles published in 2020 attracted substantially more readers than articles published earlier.

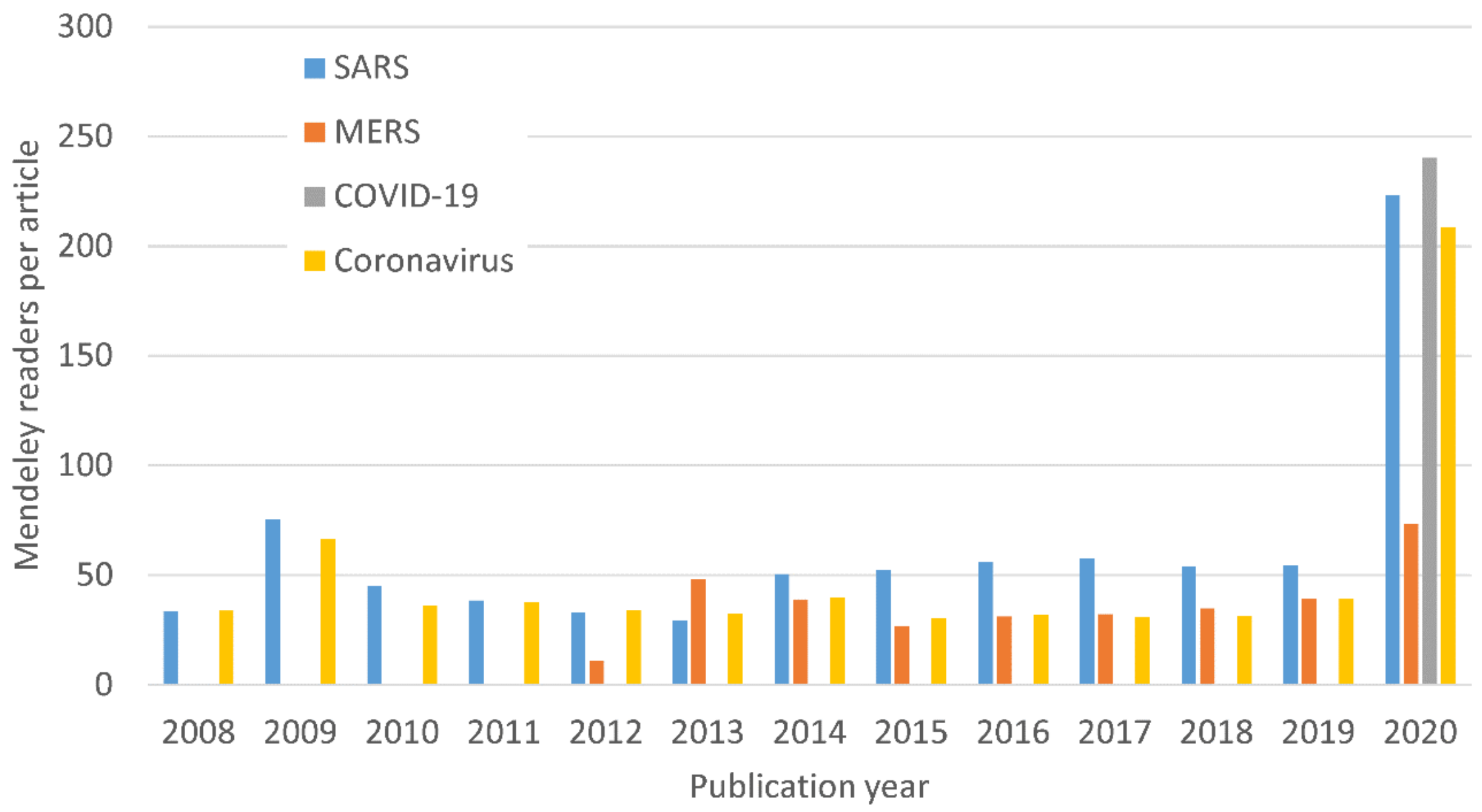

Figure 2. The average (geometric mean) number of Mendeley readers per article on 30 May 2020 for journal articles matching Dimensions "coronavirus" queries and with a title containing "SARS", "MERS" or "Coronavirus", and with a Mendeley record on 21 March 2020. "SARS" documents exclude those with "SARS-" in the title.

For all datasets, the articles published in 2020 have already attracted far more readers, on average, than articles published before (Figures 3, 4, 5, 6). For MERS (Figure 4), articles written when the disease was first identified tend to have attracted more readers, presumably due to their use in follow-up research.

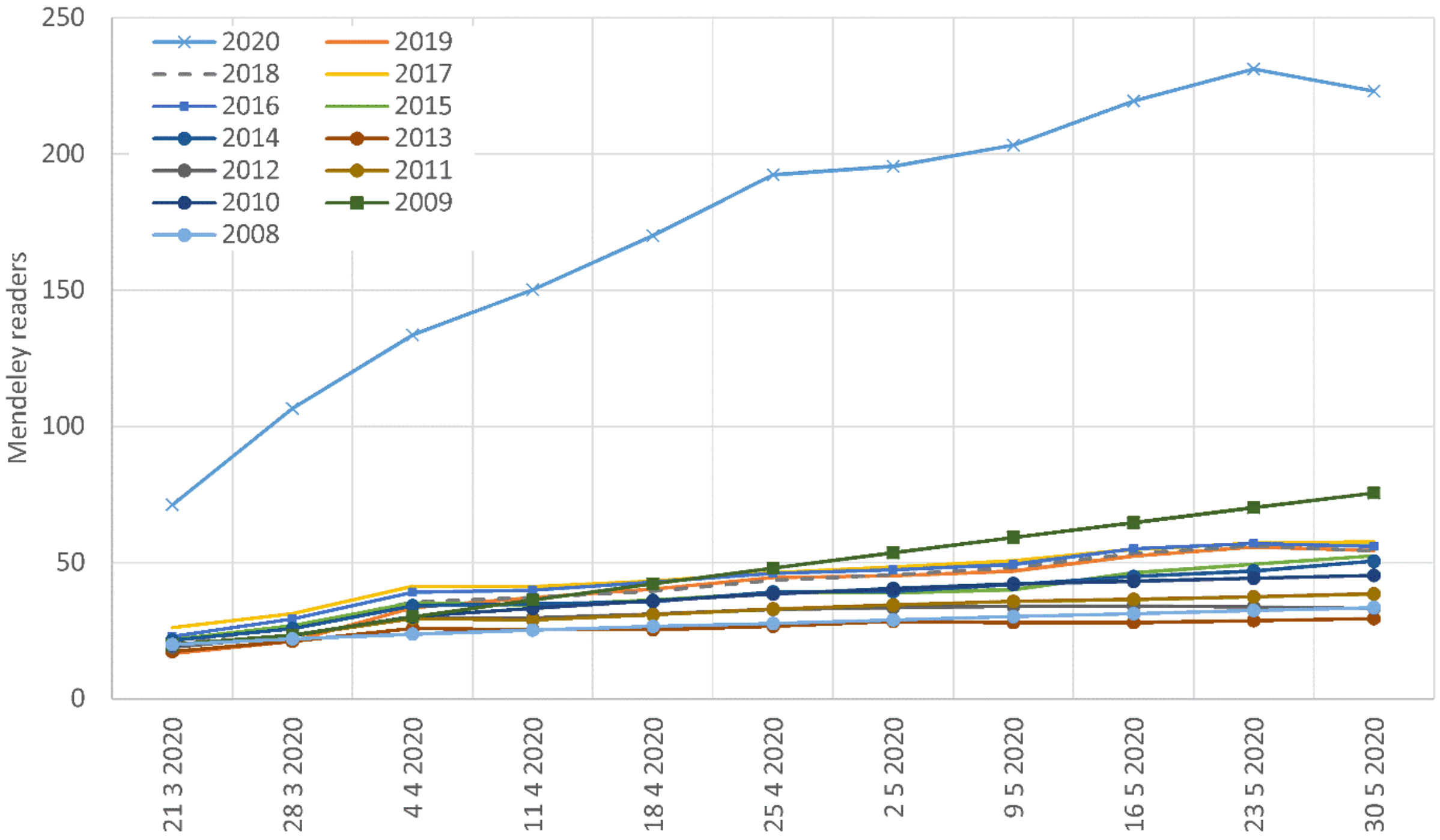

Figure 3. The weekly average (geometric mean) number of Mendeley readers per article on 30 May 2020 for journal articles matching Dimensions "coronavirus" queries and with a title containing "SARS" or a synonym, and with a Mendeley record on 21 March 2020. Data for individual years are in the supplementary information.

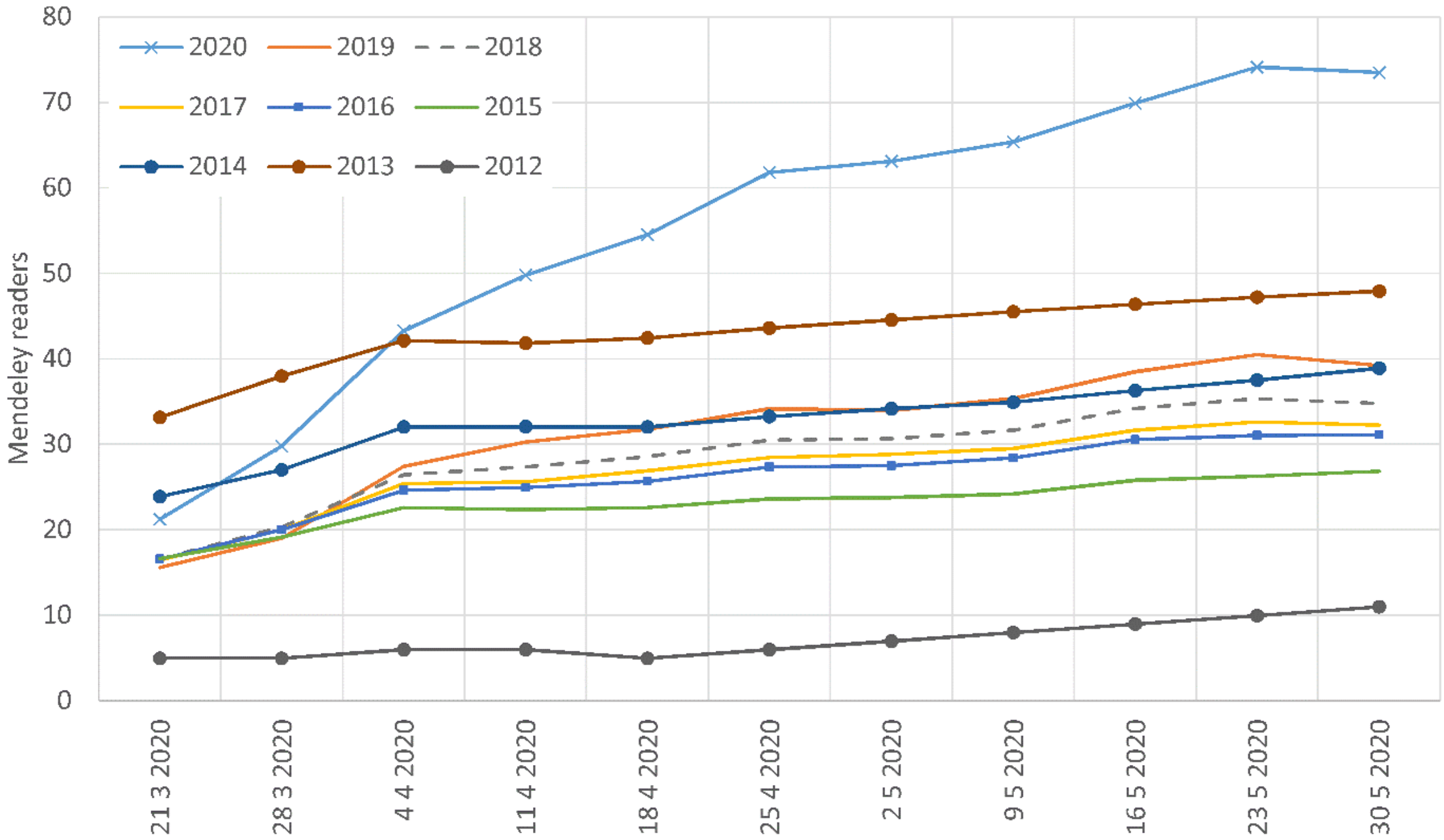

Figure 4. The weekly average (geometric mean) number of Mendeley readers per article on 30 May 2020 for journal articles matching Dimensions "coronavirus" queries and with a title containing "MERS" or a synonym, and with a Mendeley record on 21 March 2020. Data for individual years are in the supplementary information.

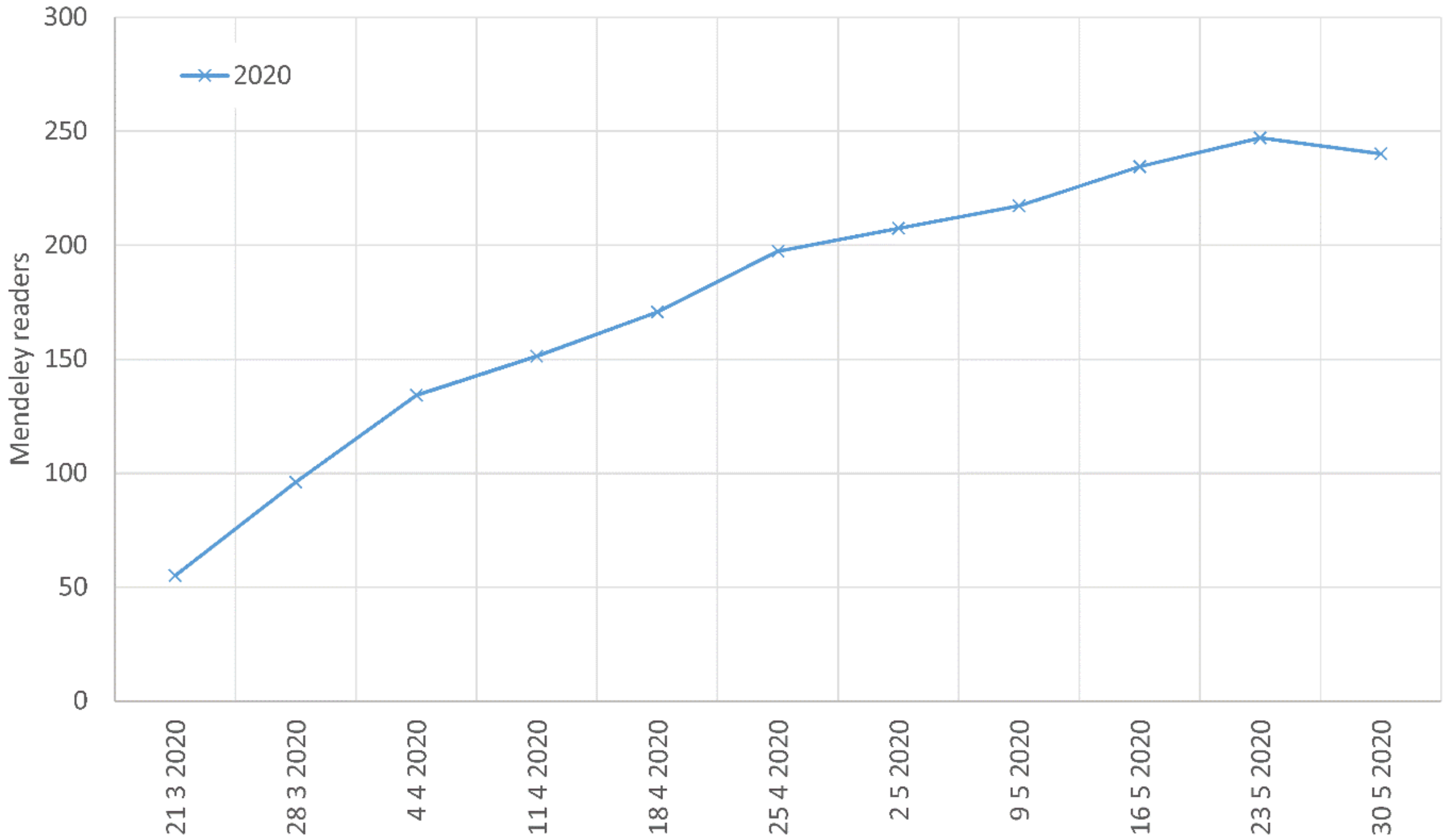

Figure 5. The weekly average (geometric mean) number of Mendeley readers per article on 30 May 2020 for journal articles matching Dimensions "coronavirus" queries and with a title containing "**COVID-19**" or a synonym, and with a Mendeley record on 21 March 2020. Data for individual years are in the supplementary information.

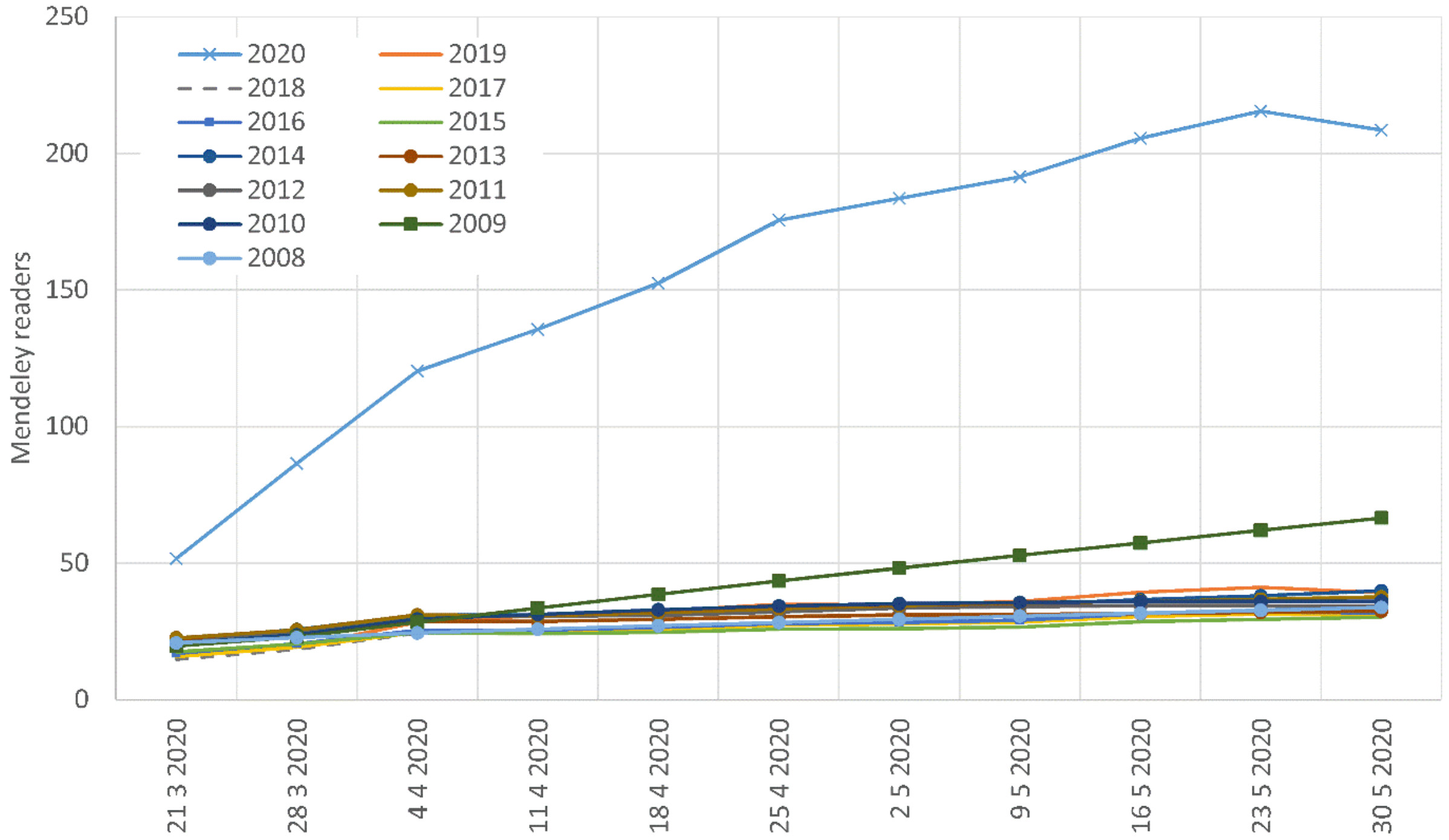

Figure 6. The weekly average (geometric mean) number of Mendeley readers per article on 30 May 2020 for journal articles matching Dimensions "coronavirus" queries and with a title containing "coronavirus" or another human coronavirus name, and with a Mendeley record on 21 March 2020. Data for individual years are in the supplementary information.

The percentage increase in Mendeley readers over a two-month period was calculated for each data set and year (Figure 7). Multiplying by 6 would give an estimated 12 month (annual) increase in Mendeley readers, if the rate was constant over the year. For reference, a 17% increase over two months would equate to an annual increase of over 100%. For almost all datasets and years, the two-month increase was above 17%, indicating an over 100% annual increase in Mendeley readers. This figure can be benchmarked against expected increases in Mendeley readers, based on prior information about the rate at which citations increase.

Citations can expect to continue to accumulate in the long term, so new citations for old articles are to be expected. For example, the cited half-life of the most common source, the Journal of Virology, is 8.9 years according to the 2019 Clarivate Journal Citation Reports, so the typical virology article should have attracted half of its final citation count after nine years. Nevertheless, in the life sciences, the annual number of citations that a paper attracts seems to peak after two years, then gradually decrease (Adams, 2005; this is an old reference but the average age of cited biomedical literature has been approximately constant since the 1950s: Larivière, Archambault, & Gingras, 2008) and in biology the peak for more cited papers published in 1990 is at 4 years, with the peak for less cited papers from 1990 being at 2 years (Parolo, Pan, Ghosh, Huberman, Kaski, & Fortunato, 2015), consistent with an overall average of 3 years to the citation peak. Thus, after three years, the annual percentage increase in citations should be substantially below 100% and after four years it should be below 50%. Since Mendely readers accumulate a year earlier than citations (Maflahi & Thelwall, 2018; Thelwall, 2017), the annual Mendeley reader count percentage increases should be substantially below 100% after two years and below 50% after three years. Thus, SARS, MERS and coronavirus research for every year from 2008 to 2017 has attracted an abnormally large increase in academic attention during April and May 2020. This is especially the case for SARS research. It is to be expected that the rate of increase is highest for the newest articles, with two anomalous exceptions (2009 and MERS in 2012 – only one paper).

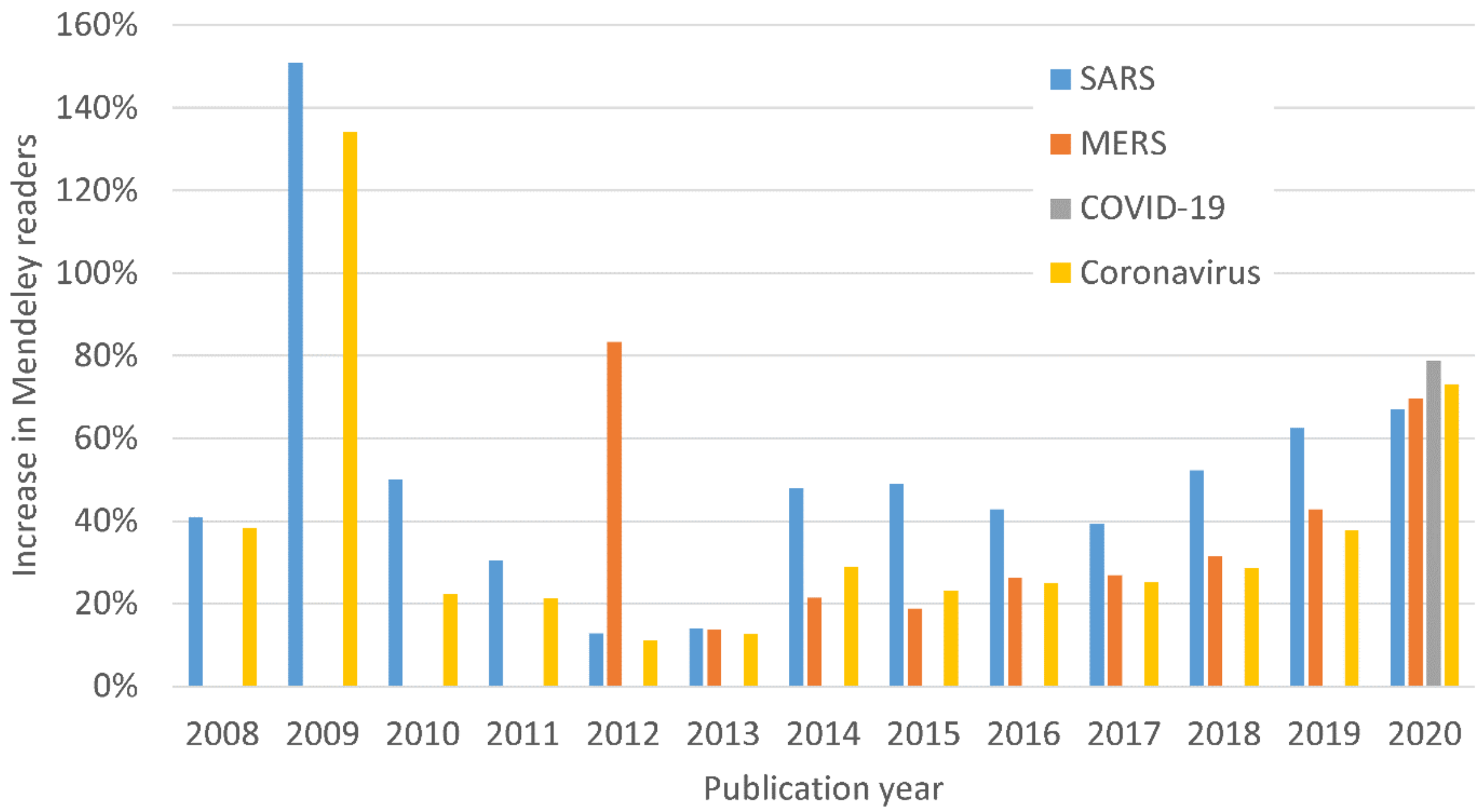

Figure 7. The percentage increase in average (geometric mean) Mendeley readers per article on 30 May 2020 compared to 4 April 2020 for journal articles matching Dimensions "coronavirus" queries and with a Mendeley record on 21 March 2020.

The reason for the high readership counts for SARS papers from 2020, compared to SARS papers from previous years (Figure 3) can be deduced by reading their titles (Table 1). The top 17 SARS papers from 2020 also mentioned COVID-19 in their titles. Thus, the high readership rate for 2020 is probably due to SARS being mentioned in the context of its implications for COVID-19.

Table 1. The 25 SARS papers published in 2020 and found by Dimensions by 21 March 2020. Mendeley reader counts are from 30 May 2020.

| **Title*** | **Readers** |
|---|---|
| Receptor recognition by the **novel coronavirus from Wuhan**: an analysis based on decade-long structural studies of SARS coronavirus. | 2235 |
| Immune responses in **COVID-19** and potential vaccines: Lessons learned from SARS and MERS epidemic. | 1555 |
| The reproductive number of **COVID-19** is higher compared to SARS coronavirus | 1493 |
| Potent binding of **2019 novel coronavirus** spike protein by a SARS coronavirus-specific human monoclonal antibody | 1007 |
| The deadly coronaviruses: The 2003 SARS pandemic and the **2020 novel coronavirus** epidemic in China | 897 |
| A systematic review of lopinavir therapy for SARS coronavirus and MERS coronavirus-A possible reference for **coronavirus disease-19** treatment option | 750 |
| From SARS to **COVID-19:** A previously unknown SARS-CoV-2 virus of pandemic potential infecting humans - Call for a One Health approach | 715 |
| Coronavirus **covid-19** has killed more people than SARS and MERS combined, despite lower case fatality rate | 579 |
| Does **SARS-CoV-2** has a longer incubation period than SARS and MERS? | 539 |
| Potential factors influencing repeated SARS outbreaks in China | 478 |
| Radiology perspective of coronavirus disease 2019 (**COVID-19**): Lessons from Severe Acute Respiratory Syndrome and Middle East Respiratory Syndrome. | 476 |
| SARS to novel coronavirus - old lessons and new lessons | 473 |
| Structural, glycosylation and antigenic variation between 2019 novel coronavirus (**2019-nCoV**) and SARS coronavirus (SARS-CoV) | 449 |
| Emerging threats from zoonotic coronaviruses-from SARS and MERS to **2019-nCoV** | 423 |
| **COVID-19**: Lessons from SARS and MERS | 407 |
| From SARS and MERS CoVs to **SARS-CoV-2**: Moving toward more biased codon usage in viral structural and nonstructural genes | 376 |
| China's response to a **novel coronavirus** stands in stark contrast to the 2002 SARS outbreak response | 366 |
| Gold nanoparticle-adjuvanted S protein induces a strong antigen-specific IgG response against severe acute respiratory syndrome-related coronavirus infection, but fails to induce protective antibodies and limit eosinophilic infiltration in lungs | **340** |
| Deja vu or jamais vu? How the Severe Acute Respiratory Syndrome experience influenced a Singapore radiology department's response to the coronavirus disease (**COVID-19**) Epidemic. | 333 |
| Inactivation of three emerging viruses - severe acute respiratory syndrome coronavirus, Crimean-Congo haemorrhagic fever virus and Nipah virus - in platelet concentrates by ultraviolet C light and in plasma by methylene blue plus visible light | **324** |
| Identification of potential cross-protective epitope between **2019-nCoV** and SARS virus | 277 |
| A high ATP concentration enhances the cooperative translocation of the SARS coronavirus helicase nsP13 in the unwinding of duplex RNA | **236** |
| Aerosol and surface stability of HCoV-19 (**SARS-CoV-2**) compared to SARS-CoV-1 | 171 |
| Long-term bone and lung consequences associated with hospital-acquired severe acute respiratory syndrome: a 15-year follow-up from a prospective cohort study | **84** |
| Evaluation of an octahydroisochromene scaffold used as a novel SARS 3CL protease inhibitor | **37** |

*References to COVID-19 are in bold. Readers are bold when there is no reference to COVID-19.

The reason for the high readership counts for MERS papers from 2020, compared to SARS papers from previous years (Figure 4) can again be deduced by reading their titles (Table 2). For MERS, most articles mentioning COVID-19 (including its earlier names) have more readers than most articles not mentioning it in its title, with some exceptions. Thus, again the high readership rate for MERS is partly due to research using it to illuminate COVID-19 properties, usually by comparisons or in the form of learning lessons from it. There are four exceptions in terms of highly read articles that do not mention COVID-19. Three are about treatments for MERS that mention the antiviral medication **remdesivir**, which has also been suggested elsewhere as a potential treatment for COVID-19. A review of MERS research is also widely read (604 readers). Three papers mentioning COVID-19 (284, 222 readers) are not in the top 18. One is a short editorial and the other two are short letters.

Table 2. The 60 MERS papers published in 2020 and found by Dimensions by 21 March 2020. Mendeley reader counts are from 30 May 2020.

| **Title*** | **Readers** |
|---|---|
| Immune responses in **COVID-19** and potential vaccines: Lessons learned from SARS and MERS epidemic. | 1555 |
| The SARS, MERS and novel coronavirus (**COVID-19**) epidemics, the newest and biggest global health threats: what lessons have we learned? | 967 |
| Potential maternal and infant outcomes from coronavirus 2019-nCoV (**SARS-CoV-2**) infecting pregnant women: lessons from SARS, MERS, and other human coronavirus infections | 863 |
| Overlapping and discrete aspects of the pathology and pathogenesis of the emerging human pathogenic coronaviruses SARS-CoV, MERS-CoV, and **2019-nCoV** | 776 |
| Comparative therapeutic efficacy of remdesivir and combination lopinavir, ritonavir, and interferon beta against MERS-CoV | **754** |
| A systematic review of lopinavir therapy for SARS coronavirus and MERS coronavirus-A possible reference for **coronavirus disease-19** treatment option | 750 |
| Prophylactic and therapeutic remdesivir (GS-5734) treatment in the rhesus macaque model of MERS-CoV infection | **747** |
| The antiviral compound remdesivir potently inhibits RNA-dependent RNA polymerase from Middle East respiratory syndrome coronavirus | **714** |
| Focus on Middle East respiratory syndrome coronavirus (MERS-CoV) | **604** |
| Coronavirus **covid-19** has killed more people than SARS and MERS combined, despite lower case fatality rate | 579 |
| Novel coronavirus **2019-nCoV**: prevalence, biological and clinical characteristics comparison with SARS-CoV and MERS-CoV. | 568 |
| Does **SARS-CoV-2** has a longer incubation period than SARS and MERS? | 539 |
| Asymptomatic coronavirus infection: MERS-CoV and SARS-CoV-2 (**COVID-19**) | 490 |
| Three emerging coronaviruses in two decades the story of SARS, MERS, and now **COVID-19** | 479 |
| Radiology perspective of coronavirus disease 2019 (**COVID-19**): Lessons from Severe Acute Respiratory Syndrome and Middle East Respiratory Syndrome. | 476 |
| Emerging threats from zoonotic coronaviruses-from SARS and MERS to **2019-nCoV** | 423 |
| **COVID-19** in the Shadows of MERS-CoV in the Kingdom of Saudi Arabia | 422 |
| **COVID-19**: Lessons from SARS and MERS | 407 |
| From SARS and MERS CoVs to **SARS-CoV-2**: Moving toward more biased codon usage in viral structural and nonstructural genes | 376 |
| MERS-CoV infection among healthcare workers and risk factors for death: Retrospective analysis of all laboratory-confirmed cases reported to WHO from 2012 to 2 June 2018 | **366** |
| Diagnosis of **SARS-CoV-2** infection based on CT scan vs. RT-PCR: Reflecting on experience from MERS-CoV | 365 |
| SARS-CoV, MERS-CoV and now the **2019-novel CoV**: Have we investigated enough about coronaviruses? - A bibliometric analysis | 362 |
| Characterization of novel monoclonal antibodies against MERS-coronavirus spike protein | **342** |
| Effect of isolation practice on the transmission of Middle East respiratory syndrome coronavirus among hemodialysis patients: A 2-year prospective cohort study. | **305** |
| Effectiveness for the Response to **COVID-19**: The MERS Outbreak Containment Procedures | 299 |
| A realistic two-strain model for MERS-CoV infection uncovers the high risk for epidemic propagation | **256** |
| Influence of trust on two different risk perceptions as an affective and cognitive dimension during Middle East respiratory syndrome coronavirus (MERS-CoV) outbreak in South Korea: serial cross-sectional surveys | **237** |
| Middle East Respiratory Syndrome-Corona Virus (MERS-CoV) associated stress among medical students at a university teaching hospital in Saudi Arabia | **224** |
| Origins of MERS-CoV, and lessons for **2019-nCoV** | 222 |
| Small molecule inhibitors of Middle East respiratory syndrome coronavirus fusion by targeting cavities on heptad repeat trimers. | **218** |
| Syndromic surveillance system for MERS-CoV as new early warning and identification approach | **211** |
| Cross-sectional prevalence study of MERS-CoV in local and imported dromedary camels in Saudi Arabia, 2016-2018 | **203** |
| Decoupling deISGylating and deubiquitinating activities of the MERS virus papain-like protease | **149** |
| Topological dynamics of the 2015 South Korea MERS-CoV spread-on-contact networks | **142** |
| Treatment of Middle East respiratory syndrome with a combination of lopinavir/ritonavir and interferon-b (MIRACLE trial): statistical analysis plan for a recursive two-stage group sequential randomized controlled trial | **139** |
| Middle East Respiratory Syndrome Coronavirus (MERS-CoV) - surveillance and testing in North England from 2012 to 2019 | **119** |
| MERS-CoV infection is associated with downregulation of genes encoding Th1 and Th2 cytokines/chemokines and elevated inflammatory innate immune response in the lower respiratory tract | **102** |
| Climate factors and incidence of Middle East respiratory syndrome coronavirus | **101** |
| Seroprevalence of MERS-CoV in healthy adults in western Saudi Arabia, 2011-2016 | **97** |
| Ribavirin and interferon therapy for critically ill patients with Middle East respiratory syndrome: a multicenter observational study | **96** |
| The risk factors associated with MERS-CoV patient fatality: A global survey | **83** |
| Polymorphisms in dipeptidyl peptidase 4 reduce host cell entry of Middle East respiratory syndrome coronavirus | **82** |
| Burden of Middle East respiratory syndrome coronavirus infection in Saudi Arabia | **74** |
| Pediatric Middle East Respiratory Syndrome Coronavirus (MERS-CoV) Infection - UAE | **70** |
| Effect of information disclosure policy on control of infectious disease: MERS-CoV Outbreak in South Korea | **67** |
| Seroprevalence of Middle East Respiratory Syndrome Corona Virus in dromedaries and their traders in upper Egypt. | **59** |
| Demographic variations of MERS-CoV infection among suspected and confirmed cases: an epidemiological analysis of laboratory-based data from riyadh regional laboratory | **57** |
| Characterization of the immune response of MERS-CoV vaccine candidates derived from two different vectors in mice | **57** |
| Middle East respiratory syndrome coronavirus antibodies in bactrian and hybrid camels from Dubai. | **55** |
| Infection prevention measures for surgical procedures during a Middle East Respiratory Syndrome outbreak in a tertiary care hospital in South Korea | **51** |
| Narrative review of Middle East respiratory syndrome coronavirus (MERS-CoV) infection: updates and implications for practice. | **48** |
| Genetic diversity of MERS-CoV spike protein gene in Saudi Arabia | **47** |
| Spatial association between primary Middle East respiratory syndrome coronavirus infection and exposure to dromedary camels in Saudi Arabia | **47** |
| Knowledge and attitudes towards Middle East respiratory sydrome-coronavirus (MERS-CoV) among health care workers in south-western Saudi Arabia. | **46** |
| Ultra-rapid real-time RT-PCR method for detecting Middle East Respiratory Syndrome coronavirus using a mobile PCR device, PCR1100 | **39** |
| Clinical characteristics of two human to human transmitted coronaviruses: **Corona virus disease 2019 versus** Middle East Respiratory Syndrome coronavirus. | 35 |
| Confronting the persisting threat of the Middle East respiratory syndrome to global health security | **35** |
| One Health-based control strategies for MERS-CoV | **15** |
| When the illiberal and the neoliberal meet around infectious diseases: an examination of the MERS response in South Korea | **14** |
| MERS-CoV infection is associated with downregulation of genes encoding Th1 and Th2 cytokines/chemokines and elevated inflammatory innate immune response in the lower respiratory tract | **8** |

*References to COVID-19 are in bold. Readers are bold when there is no reference to COVID-19.

## 5 Discussion

The main limitation of this study is its restriction to articles mentioning the diseases in their titles. Articles could be primarily about them without including their names in the titles, if they use an uncommon or early name variant. A second limitation is the use of Mendeley reader counts as a source of academic impact evidence. Mendeley may be less used in China and this could skew the results away from studies that were more read in China. Also, its use as an academic impact source may be misleading if users employ it for non-academic purposes (such as personal safety) during the pandemic. The use of a two-month period to assess changes is also a restriction since the rate of increase of readers might speed or slow in the rest of 2020.

The results show, apparently for the first time, that older SARS and MERS research has generated substantial new attention in April-May 2020, which is almost certainly due to new COVID-19 research. Nevertheless, SARS and MERS research from before 2020 has had far less academic impact, at least as reflected by Mendeley reader counts, than articles from 2020 that have reviewed or situated prior SARS and MERS research in the context of COVID-19. This issue does not seem to have been investigated for other groups of related diseases. Thus, whilst studies of SARS and MERS have informed COVID-19 research, this has occurred disproportionately through new articles that have explicitly made connections with COVID-19 or that have translated SARS and MERS research into implications for COVID-19.

## 6 Conclusions

The results confirm that older SARS and MERS research is proving useful for COVID-19, but also suggest that research interpreting SARS and MERS studies for COVID-19 performs a useful role in academia. In future, when new diseases emerge that are variants of known diseases, researchers may therefore need to prioritise publishing reviews of prior research targeting a new disease at its early stages as a service to scientists researching the new disease. These reviews may save valuable time by reducing the need for researchers and clinicians to rely on the source material to draw conclusions about lessons for the new disease.

**DATA AVAILABILITY**

The processed data used to produce the graphs are available in the supplementary material (https://doi.org/10.6084/m9.figshare.12442616).

## References

Adams, J. (2005). Early citation counts correlate with accumulated impact. *Scientometrics*, 63(3), 567-581.

Chahrour, M., Assi, S., Bejjani, M., Nasrallah, A. A., Salhab, H., Fares, M., & Khachfe, H. H. (2020). A bibliometric analysis of Covid-19 research activity: A call for increased output. *Cureus*, 12(3).

Colavizza, G., Costas, R., Traag, V. A., Van Eck, N. J., Van Leeuwen, T., & Waltman, L. (2020). A scientometric overview of CORD-19. *BioRxiv*.

Danesh, F., & GhaviDel, S. (2020). Coronavirus: Scientometrics of 50 Years of global scientific productions. *Iranian Journal of Medical Microbiology*, 14(1), 1-16.

Dehghanbanadaki, H., Seif, F., Vahidi, Y., Razi, F., Hashemi, E., Khoshmirsafa, M., & Aazami, H. (2020). Bibliometric analysis of global scientific research on Coronavirus (COVID-19). *Medical Journal of The Islamic Republic of Iran (MJIRI)*, 34(1), 354-362.

Gómez-Ríos, D., López-Agudelo, V. A., & Ramírez-Malule, H. (2020). Repurposing antivirals as potential treatments for SARS-CoV-2: From SARS to COVID-19. *Journal of Applied Pharmaceutical Science*, 10(5), 1-9.

Haghani, M., Bliemer, M. C., Goerlandt, F., & Li, J. (2020). The scientific literature on Coronaviruses, COVID-19 and its associated safety-related research dimensions: A scientometric analysis and scoping review. *Safety Science*, 104806. https://doi.org/10.1016/j.ssci.2020.104806

Hamidah, I., Sriyono, S., & Hudha, M. N. (2020). A bibliometric analysis of Covid-19 research using VOSviewer. *Indonesian Journal of Science and Technology*, 5(2), 34-41.

Hossain, M. M., (2020). Current status of global research on novel coronavirus disease (COVID-19): a bibliometric analysis and knowledge mapping [version 1; peer review: awaiting peer review]. *F1000Research*, 9, 374.

Hu, Y. J., Chen, M. M., Wang, Q., Zhu, Y., Wang, B., Li, S. F., & Hu, Y. H. (2020). From SARS to COVID-19: A bibliometric study on emerging infectious diseases with natural language processing technologies. https://assets.researchsquare.com/files/rs-25354/v1/bmcarticle.pdf

Kagan, D., Moran-Gilad, J., & Fire, M. (2020). Scientometric trends for coronaviruses and other emerging viral infections. *BioRxiv*.

Kambhampati, S. B., Vaishya, R., & Vaish, A. (2020). Unprecedented surge in publications related to COVID-19 in the first three months of pandemic: A bibliometric analytic report. *Journal of Clinical Orthopaedics and Trauma*. https://doi.org/10.1016/j.jcot.2020.04.030

Kousha, K. & Thelwall, M. (2020). COVID-19 publications: Database coverage, citations, readers, tweets, news, Facebook walls, Reddit posts. *Quantitative Science Studies*.

Larivière, V., Archambault, É., & Gingras, Y. (2008). Long-term variations in the aging of scientific literature: From exponential growth to steady-state science (1900–2004). *Journal of the American Society for Information Science and Technology*, 59(2), 288-296.

Latif, S., Usman, M., Manzoor, S., Iqbal, W., Qadir, J., Tyson, G., & Crowcroft, J. (2020). Leveraging data science to combat COVID-19: A comprehensive review. https://www.techrxiv.org/articles/Leveraging_Data_Science_To_Combat_COVID-19_A_Comprehensive_Review/12212516/1

Lou, J., Tian, S. J., Niu, S. M., Kang, X. Q., Lian, H. X., Zhang, L. X., & Zhang, J. J. (2020). Coronavirus disease 2019: a bibliometric analysis and review. *European Review of Medical and Pharmacological Science*, 24(6), 3411-21.

Maflahi, N, & Thelwall, M. (2018). How quickly do publications get read? The evolution of Mendeley reader counts for new articles. *Journal of the Association for Information Science and Technology*, 69(1), 158–167.

Mohammadi, E., Thelwall, M., Haustein, S., & Larivière, V. (2015). Who reads research articles? An altmetrics analysis of Mendeley user categories. *Journal of the Association for Information Science and Technology*, 66(9), 1832-1846.

Mohammadi, E., Thelwall, M., & Kousha, K. (2016). Can Mendeley bookmarks reflect readership? A survey of user motivations. *Journal of the Association for Information Science and Technology*, 67(5), 1198-1209.

Parolo, P. D. B., Pan, R. K., Ghosh, R., Huberman, B. A., Kaski, K., & Fortunato, S. (2015). Attention decay in science. *Journal of Informetrics*, 9(4), 734-745.

Tao, Z., Zhou, S., Yao, R., Wen, K., Da, W., Meng, Y., & Tao, L. (2020). COVID-19 will stimulate a new coronavirus research breakthrough: a 20-year bibliometric analysis. *Annals of Translational Medicine*, 8(8), 528. https://doi.org/10.21037/atm.2020.04.26

Thelwall, M. (2017a). Are Mendeley reader counts high enough for research evaluations when articles are published? *Aslib Journal of Information Management*, 69(2), 174-183.

Thelwall, M. (2017b). Are Mendeley reader counts useful impact indicators in all fields? *Scientometrics*, 113(3), 1721–1731.

Thelwall, M. (2018). Early Mendeley readers correlate with later citation counts. *Scientometrics*, 115(3), 1231–1240.

Torres-Salinas, D. (2020). Ritmo de crecimiento diario de la producción científica sobre Covid-19. Análisis en bases de datos y repositorios en acceso abierto. *El Profesional de la Información*, 29(2), e290215. https://doi.org/10.3145/epi.2020.mar.15

Zahedi, Z., Costas, R., & Wouters, P. (2014). How well developed are altmetrics? A cross-disciplinary analysis of the presence of 'alternative metrics' in scientific publications. *Scientometrics*, 101(2), 1491-1513.

Zahedi, Z., Haustein, S., & Bowman, T. (2014). Exploring data quality and retrieval strategies for Mendeley reader counts. In *SIG/MET Workshop, ASIS&T 2014 Annual Meeting*, Seattle. http://www. asis.org/SIG/SIGMET/data/uploads/sigmet2014/zahedi.pdf

Zhang, L., Zhao, W. J., Sun, B. B., Huang, Y., & Glänzel, W. (2020). How scientific research reacts to international public health emergencies: a global analysis of response patterns. *Scientometrics*.